\documentclass[twocolumn,showpacs,preprintnumbers,amsmath,amssymb]{revtex4}
\usepackage{bm}
\usepackage{graphicx}	

\begin{document}
\title{         
High density quark matter
in the NJL model
with dimensional vs. cut-off regularization
}
\author{
T. Fujihara, D. Kimura
}
\affiliation{
Department of Physics, Hiroshima University,
Higashi-Hiroshima, Hiroshima
739-8526, JAPAN}
\author{
T. Inagaki
}
\affiliation{
Information Media Center, Hiroshima University,
Higashi-Hiroshima, Hiroshima
739-8521, JAPAN}
\author{
A. Kvinikhidze
}
\affiliation{
A. Razmadze Mathematical Institute of Georgian Academy of Sciences, 
M. Alexidze Str. 1, 380093 Tbilisi, Georgia
}

\date{\today}

\begin{abstract}
We investigate  color superconducting phase at high density in the 
extended Nambu--Jona-Lasinio  model for the two flavor quarks. 
Because of the non-renormalizability of the model,
physical observables may depend on the regularization procedure, that is 
why we apply two types of regularization, the  cut-off
 and the dimensional one to evaluate the phase 
structure, the equation of state and the relationship between the mass and 
the radius of a dense star. To obtain the phase structure we evaluate the 
minimum of the effective potential at finite temperature and chemical 
potential. The stress tensor is calculated to derive the equation of 
state. Solving the Tolman-Oppenheimer-Volkoff equation, we show the 
relationship between the mass and the radius of a dense star. 
The dependence on the regularization is found not to be small, 
interestingly, dimensional regularization predicts color
superconductivity phase at rather large values of $\mu$ 
(in agreement with perturbative QCD in contrast to the cut-off 
regularization), in the larger temperature interval, the existence 
of heavier and larger quark stars.
\end{abstract}


\maketitle

\section{Introduction}
Variety of new phases of the matter consisting of quarks and gluons is expected in 
the high density QCD. One of them is the broken color 
$SU_c(3)$ symmetry phase where the color superconductivity takes place, 
Refs. [1-4] 
Quark-quark interaction is attractive in the color 
anti-symmetric $\bar{3}$ channel. This force destabilizes the filed 
Fermi sea leading to the Cooper instability. Then a composite operator 
constructed from a color anti-triplet diquark pair develops a 
non-vanishing expectation value. For the case of the two flavor quarks the diquark 
condensation breaks the color $SU_c(3)$ symmetry down to the $SU_c(2)$.
This broken symmetry phase is called two-flavor color 
superconducting (2SC) phase.
Because of the high degeneracy of states near the Fermi surface, such a 
symmetry breakdown behavior is a subject of a non-perturbative description in QCD. 

To investigate non-perturbative QCD effects we are working in the low energy
effective theory, namely in the Nambu--Jona-Lasinio (NJL) model in which 
the chiral symmetry is broken down dynamically \cite{NJL}. 
The model is useful to evaluate low-energy phenomena in  
hadronic phase of QCD.
The NJL model is extended in a simple way to include the attraction 
in the $\bar{3}$ $qq$ channel.
It is based on the point-like four-fermion interactions between quarks. 
Since the four-fermion interaction is the dimension 6 
operator, the model is non-renormalizable in four space-time dimensions, therefore
some regularization is needed to obtain finite results.

Most of analysis have been using momentum cut-off regularization, where
the cut-off scale  
  is determined phenomenologically. 
Unfortunately this cut-off scale often breaks some symmetries of the model. 
Moreover the critical chemical potential where the color 
superconductivity takes place is of the order of the cut-off
scale (ultraviolet cut-off may even hit the Fermi sea cut-off). In such a 
situation, it is expected that the regularization 
procedure has a non-negligible effect on the analysis of the 
color superconductivity.
That is why in the present paper we analyze the extended NJL model by 
using the dimensional regularization as well.
In the dimensional regularization the space-time dimension is 
analytically continued to less than four. 

Some high density states are observed in astrophysical objects.
In the core of dense stars, like neutron stars interior, quark stars 
and so on, the color superconducting phase is expected to take place. 
Much attention has been paid to such dense stars to find an
evidence of the color superconductivity \cite{Pro,Alfo:07}.
Characteristics of the dense stars such as the mass and the radius
are of great interest. 
A constraint on the mass and the radius of the star can be found by solving 
the Tolman-Oppenheimer-Volkoff (TOV) equation \cite{Oppenheimer:1939ne}. 
This solution depends on the equation of state (EoS), i.e. on the relationship 
between the energy density and the pressure, of the quark matter inside 
the stars. 
It was pointed out that the existence of the color superconducting 
phase may decrease the minimum of the radius of such dense stars
\cite{Pro,Alfo:07}, [9-29].
The minimal size of the stars is analyzed precisely. A finite contribution 
from the color neutrality and the $\beta$ equilibrium are analyzed
in [11-15]. 
Nucleon and vector meson
contributions are calculated in [16-19]. 
Rotating compact stars are  considered in \cite{Panda:2006}. 

In the paper the extended NJL model is regarded as a low energy 
effective theory of QCD. We consider two flavor  quarks and 
investigate properties of compact stars in the 2SC phase.
In Sec.~II we introduce the model Lagrangian and the fermion propagator 
in color superconducting phase.
In Sec.~III the phase structure of the model is evaluated at finite 
temperature and chemical potential. 
We numerically calculate expectation values of the composite operators 
$\langle\bar{\psi}\psi\rangle$ and 
$\langle\overline{\psi^c}\psi\rangle$ and 
the quark number susceptibility. 
In Sec.~IV we evaluate the energy-momentum tensor and determine the EoS
. In Sec.~V we solve the TOV equation with the account of 
the obtained EoS and show the relationship between the radius and the 
mass of a dense star in the color superconducting phase. Finally we give 
concluding remarks. 

\section{Extended NJL model}

In a state with large chemical potential the $\bar{3}$ diquark channel interaction
plays essential role in creation of a color Cooper pair of quarks. 
To study color superconductivity one can extend the NJL model to
include the $\bar{3}$ diquark interactions explicitly.
The temperature, $T$, and the chemical potential, $\mu$, is 
introduced to the theory via the imaginary time formalism.
The model is defined by the Lagrangian density 
\begin{eqnarray}
 \mathcal{L} &=& \bar{\psi}(i \partial\!\!\!/ -m-i\mu\gamma_4) \psi
 \nonumber \\
 && + G_S\{(\bar{\psi}\psi)^2
 + (\bar{\psi}_{aj}i\gamma_5 {\bm{\tau}}_{jk} \psi_{ak})^2\}  
\label{lag} \\ 
&&
 + G_D(i\overline{\psi^c}_{aj}\varepsilon_{jk}
 \epsilon^b_{ad} \gamma_5\psi_{d k})
 (i\bar{\psi}_{f l}\varepsilon_{lm}\epsilon^b_{f g} 
 \gamma_5\psi^c{}_{g m}) ,\nonumber 
\end{eqnarray}
where the indices $a,b,d,e,f,g$ and 
$j,k,l,m$ denote the colors(1,2,3) and flavors($u,d$) of the quarks, 
$m$ is the quark mass matrix, diag($m_u,m_d$), 
${\bm{\tau}}_{jk}$ represents the isospin Pauli matrices, $\psi^c$ 
is the charge conjugate of the field $\psi$.  $G_S$ 
and $G_D$ are the effective coupling constants for the $q\bar q$ scalar 
and the diquark channel respectively. 
The third line in Eq.~(\ref{lag}) exposes the attractive force 
in the $\bar{3}$ channel of the quark-quark interaction.

For practical calculations it is  convenient to introduce the 
auxiliary fields: scalar $\sigma$,
pseudo-scalar ${\bm{\pi}}$ and diquark $\Delta^b$,
to write down the Lagrangian density as 
\begin{equation}
 \mathcal{L}_{\rm aux} =
  \frac{1}{2}\bar{\Psi}G^{-1}\Psi 
  -\frac{1}{4G_S}(\sigma^2+\bm{\pi}^2)
  -\frac{1}{4G_D}|\Delta^b|^2 ,
\label{eq:Laux_NGform}
\end{equation}
where  the Nambu-Gor'kov representation \cite{NG} is used,
\begin{eqnarray}
 \Psi &=& \left(\begin{array}{c}
           \psi \\ 
           \psi^c
                \end{array}
          \right) ,
\end{eqnarray}
and $G$ represents the quark propagator,
\begin{eqnarray}
G^{-1} \equiv \hspace{70mm}  \\
\left(
   \begin{array}{c}
    i\partial\!\!\!/-m-i\mu\gamma_4-\sigma -i\gamma_5
     \bm{\pi}\cdot\bm{\tau} \qquad
     -i\Delta^a\varepsilon\epsilon^a\gamma_5 \\
    -i{\Delta^b}^*\varepsilon\epsilon^b\gamma_5 \qquad
     i\partial\!\!\!/-m+i\mu\gamma_4-\sigma - i\gamma_5
     \bm{\pi}\cdot\bm{\tau}^T
   \end{array}
  \right) . \nonumber
\label{g:inv}
\end{eqnarray}
From the equation of motion for the auxiliary fields we get the
following correspondence,
\begin{eqnarray}
\sigma&\sim& -G_S \bar{\psi}\psi , \\
{\bm{\pi}} &\sim& -G_S \bar{\psi}i\gamma_5{\bm{\tau}}\psi , \\
\Delta^b &\sim& -G_D i\overline{\psi^c}_{aj}\varepsilon_{jk}\epsilon^b_{ad}
\gamma_5\psi_{d k} .
\end{eqnarray}

The generating functional for the Lagrangian density (\ref{eq:Laux_NGform}) 
is given by
\begin{eqnarray}
 Z&=& \int \mathcal{D}(\Psi,\bar{\Psi},\Delta^b,{\Delta^b}^*,\sigma,\bm{\pi})
  e^{\int d^4x\left(\mathcal{L}_{\rm aux}
     -\frac{1}{2}\bar{\Psi}\cdot J_{\bar{\Psi}}
     -\frac{1}{2}J_{\Psi}\cdot\Psi\right)}\nonumber \\
 &=& {\cal N}\int\mathcal{D}(\Delta^b,{\Delta^b}^*,\sigma,\bm{\pi})
  \det{}^{1/2} G^{-1} \nonumber \\
 &&
  \times
  \exp\left[- \int d^4 x\ \left(\frac{\sigma^2+\bm{\pi}^2}{4G_S}
  +\frac{|\Delta^b|^2}{4G_D} \right)\right. \nonumber \\
 && -\left.\frac{1}{2} \int d^4x d^4y\ J_{\Psi}(x)
   G(x-y) J_{\bar{\Psi}}(y)
  \right] , 
\label{gen}
\end{eqnarray}
where $\cal{N}$ is a constant and the sources are
\begin{eqnarray}
     J_{\bar{\Psi}}
        \equiv\left(
          \begin{array}{c}
           J_{\bar{\psi}} \\
           J_{\overline{\psi^c}}
          \end{array}\right)
  ,\ \  J_{\Psi}\equiv(J_{\psi},  J_{\psi^c}) .
\end{eqnarray}
The quark mass affects the scalar $\sigma$ channel \cite{Fuji-07}, 
we set $\bm{\pi}= {\bm 0}$.  $SU_c(3)$ symmetry allows one to set 
$\Delta^{1,2}=0$ and $\Delta\equiv\Delta^3$. 
Here we study the phase structure for the case of the two flavor quarks.

\section{Phase Structure}
The phase structure is determined by the effective potential.
We set $J_{\Psi}=J_{\bar\Psi}=0$ in the generating functional (\ref{gen})
to obtain the effective potential, 
\begin{eqnarray}
V_{\rm eff}(\sigma, \Delta)
 &=&\frac{\sigma^2}{4G_S}+\frac{\Delta^2}{4G_D}
  - \frac{1}{ 2 \beta \Omega }\ln\det\left(G^{-1}\right)\!.
\end{eqnarray}
where $\Omega$ denotes the volume of the system, 
$\Omega \equiv \int d^{D-1} x$.
Due to the space-time translational invariance
 the expectation values of $\sigma$ and $\Delta$
are independent of the space-time coordinates. 
Assuming the isospin symmetry, $m_u=m_d=m$ and
$\mu_u=\mu_d=\mu$, the effective potential reads
\begin{eqnarray}
 V_{\rm eff}(\sigma, \Delta)
  &=&
  \frac{1}{4G_S}\sigma^2+\frac{1}{4G_D}\Delta^2
\nonumber \\ &&
  -{\rm tr}{\bf 1}_{\rm spinor}\int\frac{d^{D-1}p}{(2\pi)^{D-1}}
  \left[
   E_++E_- + E   
  \right. \nonumber \\
 &&
   +2\frac{1}{\beta}\ln \left\{(1+e^{-\beta E_+})(1+e^{-\beta E_-}) \right\}
\nonumber \\
  && \left. 
   +\frac{1}{\beta}\ln \left\{(1+e^{-\beta\xi_+})(1+e^{-\beta\xi_-}) \right\}
  \right],
\label{veff:af}
\end{eqnarray}
where ${\rm tr}{\bf 1}_{\rm spinor}=2^{D/2}$, $\beta=1/T$ and 
\begin{equation}
  E \equiv \sqrt{\bm{p}^2 + (\sigma+m)^2},\
  \xi_{\pm} \equiv E \pm \mu,\
  E_{\pm} \equiv \sqrt{ \xi_{\pm}^2 + \Delta^2 }.
\end{equation}
Since the third term of the right hand side of Eq.~(\ref{veff:af})
involves a divergent integral,
we use the dimensional regularization ($2<D<4$) to regularize it
\cite{IKM,Inag-08}. After  integration over the angle
variables, we have
\begin{eqnarray}
 V_{\rm eff}(\sigma,\Delta)&=&\frac{\sigma^2}{4G_S}+\frac{\Delta^2}{4G_D}
\label{veff:ang}
 - \frac{\tilde{A}}{\Gamma\left(\frac{D-1}{2}\right)}
\\ && \times
  \int^\infty_0
  dp \ p^{D-2}
  \left[
   E_+ + E_- + E  \right. \nonumber \\
 &&
   +2\frac{1}{\beta}\ln(1+e^{-\beta E_+})
   +2\frac{1}{\beta}\ln(1+e^{-\beta E_-}) \nonumber \\
 &&
  \left.
   +\frac{1}{\beta}\ln(1+e^{-\beta \xi_+})
   +\frac{1}{\beta}\ln(1+e^{-\beta \xi_-})
  \right], \nonumber
\end{eqnarray}
with $\tilde{A} \equiv 4\sqrt{\pi} / (2\pi)^{D/2}$. For the cut-off
regularization, the third term of the right hand side is obtained by the replacement
\begin{eqnarray}
\frac{\tilde{A}}{\Gamma\left(\frac{D-1}{2}\right)}
\int^\infty_0 dp \ p^{D-2}
\to \frac{2}{\pi^2}\int^\Lambda_0 dp \ p^{2} ,
\label{dim-cut}
\end{eqnarray}
where $\Lambda$ is a cut-off scale. 
The first three terms of the integral in Eq.~(\ref{veff:ang}) 
are naively divergent. The divergent part of the 
Eq.~(\ref{veff:ang})  in the dimensional
regularization reads
\begin{eqnarray}
&& \hspace{-5mm}
\int^\infty_0 dp \ p^{D-2}
  \left( E_+ + E_- + E  \right)  
\nonumber \\
&=&
\int^\infty_0 dp \ p^{D-2}
  \left( E_+ + E_- - 2 \sqrt{E^2 + \Delta^2}  \right) 
\nonumber \\
&&- \frac{\Gamma(-\frac{D}{2})\Gamma(\frac{D-1}{2})}
{4\sqrt{\pi}}
\left[2\{ (\sigma+m)^2+\Delta^2\}^{D/2}
\right. 
\nonumber \\
&&+\left. \{ (\sigma+m)^2\}^{D/2} \right] . 
\end{eqnarray}

The following renormalization conditions are used:
\begin{eqnarray}
 \frac{1}{2G_S^r}&=&
  \left.
   \frac{\partial^2 V_{\rm eff}}{\partial\sigma\partial\sigma}
  \right|_{\mu=T=\Delta=0,\sigma=M_0}
\nonumber \\
  &=&\frac{1}{2G_S} 
\label{cond1:ren}\\
  &&+\frac{3D(D-1)}{(2\pi)^{D/2}}\Gamma\left(-\frac{D}{2}\right)
  \left[(M_0+m)^2 \right]^{D/2-1} \nonumber
\end{eqnarray}
and
\begin{eqnarray}
 \frac{1}{2G_D^r}&=&
 \left.
  \frac{\partial^2 V_{\rm eff}}{\partial\Delta\partial\Delta}
 \right|_{\mu=T=\Delta=0,\sigma=M_0} 
\nonumber \\
 &=&
    \frac{1}{2G_D} 
\label{cond2:ren}\\
 &&+\frac{2D}{(2\pi)^{D/2}}\Gamma\left(-\frac{D}{2}\right)
    \left[(M_0+m)^2 \right]^{D/2-1} , \nonumber 
\end{eqnarray}
where $M_0$ is a renormalization scale.

The parameters of the model are fixed to reproduce
some observables at $T=\mu=0$.
Using the pion mass 
$m_\pi =138$~MeV, the pion decay constant $f_\pi = 92.4$~MeV
and the quark mass $m = 4.5$~MeV, we obtain   
$\Lambda = 720$~MeV and $G_S = 3.80\times10^{-6}~{\rm MeV}^{-2}$ for 
the cut-off regularization.
Here we use the relationship, $G_D = (3/4)G_S$, corresponding to 
the one gluon exchange interaction. 
In the chiral limit, the critical temperature becomes
170~MeV at $\mu = 0$ with these parameters. 

In the dimensional regularization one can not avoid renormalization 
which introduces an additional parameter, renormalization scale, $M_0$.
In Ref.~\cite{IKM} the scale is simply fixed at the
dynamically generated fermion mass. The similar procedure
is also adopted in \cite{Naka:92}.
Here we fix all parameters by reproducing physical observables. 
Reproduction of the critical temperature places an additional constraint 
on the parameters.
We fix the parameters for the dimensional regularization
to reproduce the pion mass, the pion decay constant, the quark mass 
and the critical temperature 170~MeV at $m\to0$ and $\mu=0$. Then we 
have $D = 2.28$, $M_0=127$~MeV,  $G_S^r = 0.603~{\rm MeV}^{2-D}$ and 
$G_D^r = (3/4)G_S^r$ for the dimensional regularization.
In Ref.~\cite{Jafa:06} the renormalization scale is fixed
by using the decay width of the $\pi^0 \to 2\gamma$. It gives
the renormalization scale near the constituent quark mass.

If the auxiliary field $\Delta\equiv\Delta^3$ develops a non-vanishing
 value, the diquark condensation takes place. It implies 
that the QCD gauge symmetry is broken down from $SU_c(3)$ to $SU_c(2)$
and the 2SC phase is realized.  
The expectation values of $\sigma$ and $\Delta$ are found by observing 
the minimum of the effective potential.
To find the minimum we numerically 
evaluate the effective potential (\ref{veff:ang}). The expectation 
values are shown as  functions of the chemical potential in Figs.~1 
and 2 near a typical temperature of dense star, $T=1$~keV, in the 
cut-off and the dimensional regularizations respectively. 
\begin{figure}[ht]
\includegraphics[width={!},height={55mm}]
{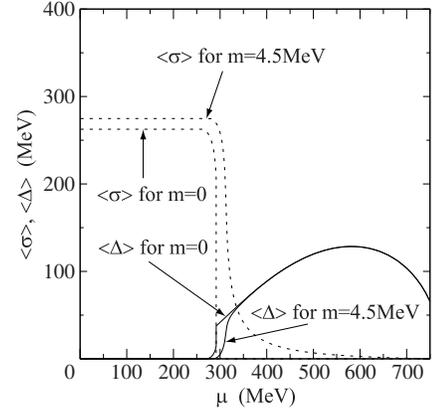}
\caption{Behavior of the expectation values of auxiliary fields 
$\sigma$ and $\Delta$ for $m=0$ and $m=4.5$~MeV at $T=1~{\rm keV}$ 
in the cut-off regularization. 
}
\end{figure}
\begin{figure}[ht]
\includegraphics[width={!},height={55mm}]
{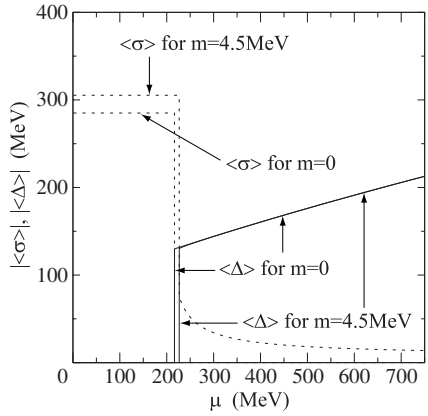}
\caption{Behavior of the expectation values of auxiliary fields 
$\sigma$ and $\Delta$ for $m=0$ and $m=4.5$~MeV at $T=1~{\rm keV}$ 
in the dimensional regularization.
}
\end{figure}
We observe  similar behavior of $\langle \sigma \rangle$ and 
$\langle \Delta \rangle$ in Figs.~1 and 2.
As is known, the current quark mass enhances the chiral symmetry
breaking and contributes to the crossover behavior, i.e. the
expectation value, $\langle \sigma \rangle$, smoothly disappears as 
the chemical potential $\mu$ increases. 
At the massless limit, the critical chemical potential for the chiral 
symmetry breaking is found to be $\mu \simeq 300$ and $220$~MeV in 
the case of the cut-off and the dimensional regularizations respectively. 
A large mass gap is observed in Fig.~2.
There is no coexistence phase with non-vanishing
$\langle \sigma \rangle$ and $\langle \Delta \rangle$ in the 
dimensional regularization. The current quark mass hardly affects 
the behavior of the expectation value $\langle \Delta \rangle$. 
The behavior of $\langle \Delta \rangle$
in Fig.~2. is quite different from that in Fig.~1. 
As is clearly seen in Fig.~2, the absolute value 
$|\langle \Delta \rangle|$ monotonically increases as a function of $\mu$.
In the cut-off regularization, $\langle \Delta \rangle$
is suppressed near the cut-off scale,
$\mu \sim \Lambda~(= 720$~MeV). 

We also calculate the quark number susceptibility $\chi_q$ 
\cite{Hatt:03,Scha:07,Cost:07,Sasa:08,Fuku:08} which is defined by 
\begin{eqnarray}
\chi_q \equiv - \frac{\partial}{\partial \mu}
 n(\langle\sigma\rangle, \langle\Delta\rangle) ,
\label{chi}
\end{eqnarray}
where $n(\langle\sigma\rangle, \langle\Delta\rangle)$ is 
the quark number density,
\begin{eqnarray}
n(\langle\sigma\rangle, \langle\Delta\rangle)
&=& \langle \psi^\dagger \psi\rangle =
-\frac{\partial }{\partial \mu}V_{\rm eff}
(\langle\sigma\rangle, \langle\Delta\rangle)
\nonumber\\
 &=&\frac{\tilde{A}}{\Gamma\left(\frac{D-1}{2}\right)}
  \int^\infty_0 dp\ p^{D-2}
\nonumber\\
 &&\times \left[
   \frac{\xi_+}{E_+}\tanh\left(\frac{\beta E_+}{2}\right) \right.
\nonumber \\
 && +\frac{1}{2}\tanh\left(\frac{\beta \xi_+}{2}\right)
    \left.  -(\mu \to -\mu)\vphantom{\frac{\xi_+}{E_+}}
  \right].
\label{n_den}
\end{eqnarray}

Figures~3 and 4 show the behavior of $\chi_q$ as a function of 
the chemical potential in the cut-off and the dimensional 
regularizations, respectively. 
\begin{figure}[ht]
\includegraphics[width={!},height={55mm}]
{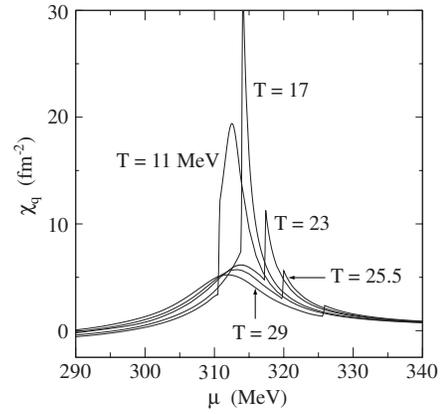}
\caption{Quark number susceptibility for 
$\mu=11, 17, 23, 25.5,$ \\
$29$
MeV in the model with cut-off regularization.}
\end{figure}
\begin{figure}[ht]
\includegraphics[width={!},height={55mm}]
{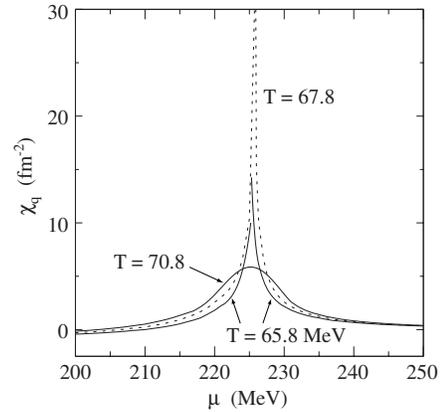}
\caption{Quark number susceptibility 
for $\mu=65.8, 67.8, 70.8$~MeV
in the model with dimensional regularization.}
\end{figure}
In Fig.~3 we show the behavior of the susceptibility $\chi_q$ 
near the phase boundary. A coexistence phase can be realized
and different critical points are observed for
$\langle \sigma \rangle$ and $\langle \Delta \rangle$ in the
cut-off regularization.
In Fig.~3 the maximum of $\chi_q$ at $\mu\simeq 312$~MeV for 
$T=29$~MeV is induced by the chiral symmetry breaking. 
Decreasing the temperature, moves this peak to the right 
until $T\sim 23$~MeV. The gap at $\mu\simeq 327$~MeV for $T=29$~MeV is 
generated by the color symmetry breaking. This gap moves to 
the left as the temperature decreases.
The susceptibility for the peak (gap)  higher than that for the 
gap (peak),  implies a crossover (the first order phase 
transition). The chemical potentials for the peak and the gap
degenerate at $T \simeq 17$~MeV, it corresponds to the critical end 
point (CEP). There is a simpler behavior of $\chi_q$ in the
dimensional regularization. 
There is no coexistence phase as is seen in Fig.~4. 
Such a picture is consistent with the Fig.~2.

The phase structure of the chiral $SU_L(2)\times SU_R(2)$ 
symmetry and 2SC in the cut-off and the dimensional regularizations
are shown
in the Figs.~5 and 6, respectively. 
We plot the boundary of the phase where $\chi_q$ has the peak or 
the gap
for $m= 4.5$~MeV and also the boundary where $\langle \sigma \rangle$ 
and/or $\langle \Delta \rangle$ disappear at the massless limit, 
$m \to 0$.
The coexistence phase induces a complex structure around the  CEP
 in the cut-off regularization. We find the tricritical 
point (TCP) at $(T, \mu)=(46.7, 281)$ and CEP at $(T, \mu)=(17.0, 314)$.
As is clearly seen in Fig.~5, the 2SC is suppressed as $\mu$ approaches 
to the cut-off scale. 
In Fig.~6 we observe TCP at $(T, \mu)=(87.2, 194)$ and CEP at (67.8, 226). 
In the dimensional regularization the 2SC gets enhanced as $\mu$ 
increases.

\begin{figure}[ht]
\includegraphics[width={!},height={55mm}]
{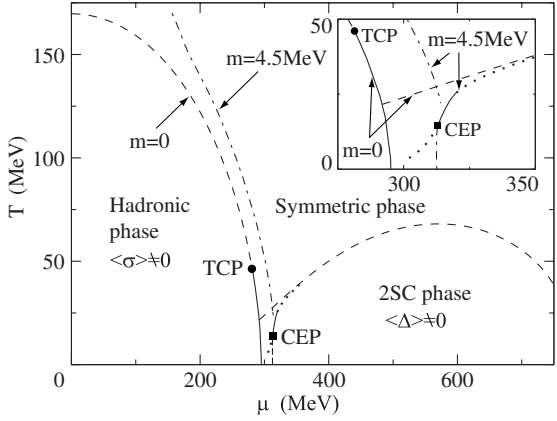}
\caption{Phase diagram for the model with the cut-off regularization.
The solid, the dashed and the dashed-dotted lines denote the first, 
the second order phase transition and the crossover, respectively.
We also plot the secondary maximum of $\chi_q$ by the dotted line.
}
\end{figure}
\begin{figure}[ht]
\includegraphics[width={!},height={55mm}]
{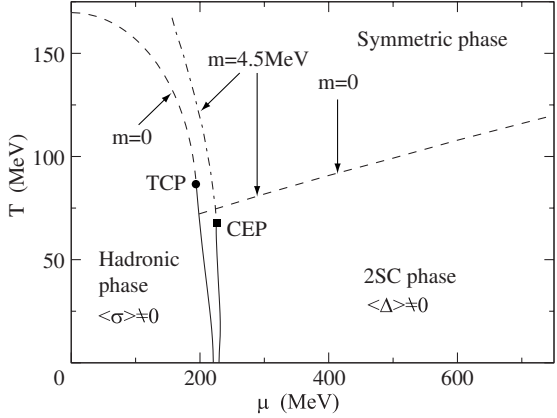}
\caption{Phase diagram for the model with the dimensional regularization.
The solid, the dashed and the dashed-dotted lines denote the first, 
the second order phase transition and the crossover, respectively.
}
\end{figure}

\section{Energy-momentum tensor}

Here we discuss the EoS in the extended NJL model. 
The energy-density and the pressure of the system are given by the 
energy-momentum tensor.
Thus we calculate the energy-momentum tensor at finite $T$ and $\mu$
in the dimensional regularization and compare the result with that 
in the cut-off regularization.  
The energy-momentum tensor is defined by 
\begin{eqnarray}
 \mathrm{T}_{\mu\nu}&=&
  \frac{\partial \mathcal{L}_{\rm aux}}{\partial(\partial^{(\mu}\psi)}
   (\partial_{\nu)}\psi)
  +(\partial_{(\nu}\bar{\psi})
   \frac{\partial\mathcal{L}_{\rm aux}}{\partial(\partial^{\mu)}\bar{\psi})}
\nonumber \\
 &&
  +\frac{\partial \mathcal{L}_{\rm aux}}{\partial(\partial^{(\mu}\psi^c)}
   (\partial_{\nu)}\psi^c)
  +(\partial_{(\nu}\overline{\psi^c})
   \frac{\partial\mathcal{L}_{\rm aux}}{\partial(\partial^{\mu)}
   \overline{\psi^c})}
\nonumber \\
 &&
  -g_{\mu\nu}\mathcal{L}_{\rm aux},
\label{st}
\end{eqnarray}
where the parenthesis in the subfix indicates symmetrization,
\begin{eqnarray*}
 A_{(\mu} B_{\nu)} = \frac12 (A_\mu B_\nu + A_\nu B_\mu).
\end{eqnarray*}
Following the imaginary time formalism, we introduce the temperature 
and the chemical potential. We perform Wick rotation and substitute
the Lagrangian density (\ref{eq:Laux_NGform}) into (\ref{st}),
\begin{eqnarray}
T_{44}&=&
 -\frac{1}{4}\bar{\psi}\gamma_{4}
        (i\partial_{4}-i\mu)\psi
 +\frac{1}{4}(i\partial_{4}+i\mu)\bar{\psi}\gamma_{4}\psi \nonumber \\
 && 
 -\frac{1}{4}\overline{\psi^c}\gamma_{4}
        (i\partial_{4}+i\mu)\psi^c
 +\frac{1}{4}
        (i\partial_{4}-i\mu)\overline{\psi^c}\gamma_{4}\psi^c\nonumber \\
 && 
 +\frac{1}{2}\bar{\psi}(i\partial\!\!\!/-i\mu\gamma_4-\sigma-m)\psi
\nonumber \\
 &&
    +\frac{1}{2}\overline{\psi^c}(i\partial\!\!\!/
    +i\mu\gamma_4-\sigma-m)\psi^c\nonumber \\
 && 
    -\frac{1}{2}\Delta(i\bar{\psi}\varepsilon\epsilon^3\gamma_5\psi^c)
    -\frac{1}{2}\Delta^*(i\overline{\psi^c}\varepsilon\epsilon^3\gamma_5\psi)
\nonumber \\
 &&
    -\frac{1}{4G_S}\sigma^2-\frac{1}{4G_D}\Delta^2,
\end{eqnarray}
and
\begin{eqnarray}
T_{ii}&=&
 -\frac{1}{4}\bar{\psi}\gamma_{i}
        i\partial_{i}\psi
 +\frac{1}{4}i\partial_{i}\bar{\psi}\gamma_{i}\psi
\nonumber \\
 && -\frac{1}{4}\overline{\psi^c}\gamma_{i}i\partial_{i}\psi^c
 +\frac{1}{4}i\partial_{i}\overline{\psi^c}\gamma_{i}\psi^c\nonumber \\
 && 
    +\frac{1}{2}\bar{\psi}(i\partial\!\!\!/-i\mu\gamma_4-\sigma-m)\psi
\nonumber \\
 &&
    +\frac{1}{2}\overline{\psi^c}(i\partial\!\!\!/
    +i\mu\gamma_4-\sigma-m)\psi^c \nonumber \\
 && 
    -\frac{1}{2}\Delta(i\bar{\psi}\varepsilon\epsilon^3\gamma_5\psi^c)
    -\frac{1}{2}\Delta^*(i\overline{\psi^c}\varepsilon\epsilon^3\gamma_5\psi)
\nonumber \\
 &&
    -\frac{1}{4G_S}\sigma^2-\frac{1}{4G_D}\Delta^2 .
\end{eqnarray}

To study phenomena in the static star, we take the thermal average of 
the energy-momentum tensor. It is given by the
expectation value in the ground state  where the equations of motion,
\begin{eqnarray}
 &&\frac{1}{2}\bar{\psi}(i\partial\!\!\!/-i\mu\gamma_4-\sigma-m)\psi
  -\frac{1}{4}\Delta^*(i\overline{\psi^c}\varepsilon\epsilon^3\gamma_5\psi)
\nonumber \\
 &&  -\frac{1}{4}\Delta(i\bar{\psi}\varepsilon\epsilon^3\gamma_5\psi^c) = 0,
\end{eqnarray}
and
\begin{eqnarray}
 &&\frac{1}{2}\overline{\psi^c}(i\partial\!\!\!/+i\mu\gamma_4-\sigma-m)\psi^c
  -\frac{1}{4}\Delta^*(i\overline{\psi^c}\varepsilon\epsilon^3\gamma_5\psi)
\nonumber \\
 &&  -\frac{1}{4}\Delta(i\bar{\psi}\varepsilon\epsilon^3\gamma_5\psi^c)
  = 0,
\end{eqnarray}
are satisfied.
Therefore the expectation values of the stress tensor elements are simplified to
\begin{eqnarray}
 \langle T_{44} \rangle &=&
 \lim_{y\to x}\frac{1}{4}\mbox{tr}\left[\gamma_{4}
        (i\partial^x_{4}-i\mu)\langle\psi(x)\bar{\psi}(y)\rangle
         \right.\nonumber \\
 &&
 -(i\partial^y_{4}+i\mu)\gamma_{4}\langle \psi(x)\bar{\psi}(y)\rangle 
\nonumber \\
 &&  
 +\gamma_{4}(i\partial^x_{4}+i\mu)\langle 
 \psi^c(x)\overline{\psi^c}(y)\rangle\nonumber \\
 && \left.
 -(i\partial^y_{4}-i\mu)\gamma_{4}\langle\psi^c(x)\overline{\psi^c}(y)
  \rangle\right]\nonumber \\
 && 
    -\frac{1}{4G_S}\langle \sigma \rangle^2
    -\frac{1}{4G_D}\langle\Delta\rangle^2 ,
\label{t44:ex}
\end{eqnarray}
and
\begin{eqnarray}
\langle T_{ii}\rangle&=&
 \lim_{y\to x}
 \frac{1}{4}\mbox{tr}\left[\gamma_{i}i\partial^x_{i}\langle
  \psi(x)\bar{\psi}(y)\rangle
 -i\partial^y_{i}\gamma_{i}\langle\psi(x)\bar{\psi}(y)\rangle \right. 
\nonumber \\
 &&  \left.
 +\gamma_{i}i\partial^x_{i}\langle\psi^c(x)\overline{\psi^c}(y)\rangle
 -i\partial^y_{i}\gamma_{i}\langle\psi^c(x)\overline{\psi^c}(y)\rangle 
\right]\nonumber \\
 && 
    -\frac{1}{4G_S}\langle \sigma\rangle^2
    -\frac{1}{4G_D}\langle\Delta\rangle^2 .
\label{tii:ex}
\end{eqnarray}
To regularize the composite operator with the quark fields 
at the same point we consider quark fields at the slightly different 
space-time points, $x$ and $y$ 
\cite{ps1,ps2,ps3}. 
 The expectation values
of the composite operators $\langle\psi(x)\bar{\psi}(y)\rangle$
and $\langle\psi^c(x)\overline{\psi^c}(y)\rangle$ are  the 
diagonal matrix elements
of the quark propagator in the Nambu-Gor'kov representation.

Substituting the fermion propagator obtained in Sec.~II into 
Eqs.~(\ref{t44:ex}) and (\ref{tii:ex}) and integrating over the 
angle variables, we obtain
\begin{eqnarray}
 \frac{\langle T_{44} \rangle}{M_0^{D-4}}
  &=&
  -\frac{1}{4G_S}\langle \sigma \rangle^2
  -\frac{1}{4G_D}\langle \Delta \rangle^2 \nonumber \\
 &&
  +\frac{\tilde{A}}{\Gamma(\frac{D-1}{2})}
  \int^\infty_0dp\ p^{D-2}
\nonumber \\
 &&\times  \left[
    \frac{E_+^2-\mu \xi_+}{E_+}\tanh\left(\frac{\beta E_+}{2}\right)
  \right.
\nonumber \\
 && \left.  
   +\frac{E}{2}\tanh\left(\frac{\beta \xi_+}{2}\right)
   +(\mu \to -\mu) \right] ,
\label{ex:t44}
\end{eqnarray}
and
\begin{eqnarray}
 \frac{\langle T_{ii} \rangle}{M_0^{D-4}}
  &=&
  - \frac{1}{4G_S}\langle \sigma \rangle^2
  - \frac{1}{4G_D}\langle \Delta \rangle^2 \nonumber \\
 &&
  - \frac{\tilde{A}}{(D-1)\Gamma\left(\frac{D-1}{2}\right)}
     \int^\infty_0dp \frac{p^{D}}{E}
\nonumber \\
 && \times
  \left[
    \frac{\xi_+}{E_+}\tanh\left(\frac{\beta E_+}{2}\right)
  \right.
\nonumber \\
 && \left.  
   +\frac{1}{2}\tanh\left(\frac{\beta \xi_+}{2}\right)
   +(\mu \to -\mu) \right] ,
\label{ex:tii}
\end{eqnarray}
where $M_0$ is the renormalization scale defined in the conditions
(\ref{cond1:ren}) and (\ref{cond2:ren}). It should be noted that
the stress tensor is renormalized to have the correct mass dimension
in the four dimensional space-time.
Taking the four dimensional limit and applying the modification 
(\ref{dim-cut}),
we numerically evaluate the integral in Eqs.~(\ref{ex:t44}) and 
(\ref{ex:tii}) in the cut-off regularization.

Since the divergent parts of the Eqs.~(\ref{ex:t44}) and (\ref{ex:tii})
do not depend on $T$ and $\mu$ explicitly,
we can evaluate the divergent parts in the limit, $\mu\to 0$ and
$\beta\to\infty$. Using the dimensional regularization, 
we evaluate  analytically the momentum integral.
Because of the general
covariance in this limit, one derives the same value for 
$\langle T_{44} \rangle$ and $\langle T_{ii} \rangle$,
\begin{eqnarray}
\frac{\langle T_{44} \rangle_0}{M_0^{D-4}}
&=& \frac{\langle T_{ii} \rangle_0}{M_0^{D-4}}
  \equiv
  -\frac{1}{4G_S}\langle \sigma \rangle^2
  -\frac{1}{4G_D}\langle \Delta \rangle^2 \nonumber \\
 &&-\frac{\Gamma(-\frac{D}{2})}{(2\pi)^{D/2}}
   \left[ 2\{(\langle\sigma\rangle+m)^2+\langle\Delta\rangle^2\}^{D/2} \right.
\nonumber \\
 &&\left.
  + \{(\langle\sigma\rangle+m)^2 \}^{D/2} \right] ,  
\label{t44_0}
\end{eqnarray}
where $\langle \sigma \rangle$ and $\langle \Delta \rangle$ depend
on $T$ and $\mu$.

Thus we obtain a finite expression for the energy-momentum tensor 
\begin{eqnarray}
\frac{\langle T_{44} \rangle}{M_0^{D-4}}
&=&
   \frac{\langle T_{44} \rangle_0}{M_0^{D-4}}
 +\frac{\tilde{A}}{\Gamma\left(\frac{D-1}{2}\right)}
 \int^\infty_0 dp\ p^{D-2}
\nonumber \\
&&\times \left[
    \frac{E_+^2-\mu \xi_+}{E_+} \left\{\tanh\left(\frac{\beta E_+}{2}\right)
     -1\right\}\right.
\nonumber \\
 && 
   +\frac{E}{2}\left\{\tanh\left(\frac{\beta \xi_+}{2}\right)
    -1\right\} +(\mu \to -\mu)
\nonumber \\
 &&+ (E_+ + E_- -2\sqrt{E^2+\langle\Delta\rangle^2})
\nonumber \\
 &&- \left.
  \mu \left( \frac{\xi_+}{E_+} - \frac{\xi_-}{E_-} \right)
  \right] ,
\label{t44}
\end{eqnarray}
and
\begin{eqnarray}
\frac{\langle T_{ii} \rangle}{M_0^{D-4}}
 &=& \frac{\langle T_{ii} \rangle_0}{M_0^{D-4}}
  -\frac{\tilde{A}}{(D-1)\Gamma(\frac{D-1}{2})}
  \int_0^\infty dp \frac{p^D}{E}
\nonumber \\
 && \times
  \left[
    \frac{\xi_+}{E_+}\left\{ \tanh\left(\frac{\beta E_+}{2}\right)
    -1 \right\} \right.
\nonumber \\
&&
   +\frac{1}{2}\left\{ \tanh\left(\frac{\beta \xi_+}{2}\right)
   -1\right\}+(\mu \to -\mu)
\nonumber \\
&& \left. 
 + \left( \frac{\xi_+}{E_+}+ \frac{\xi_-}{E_-}
   -\frac{2E}{\sqrt{E^2+\langle\Delta\rangle^2}} \right) \right] .
\label{tii}
\end{eqnarray}
\begin{figure}[ht]
 \includegraphics[width=!,height=55mm]
 {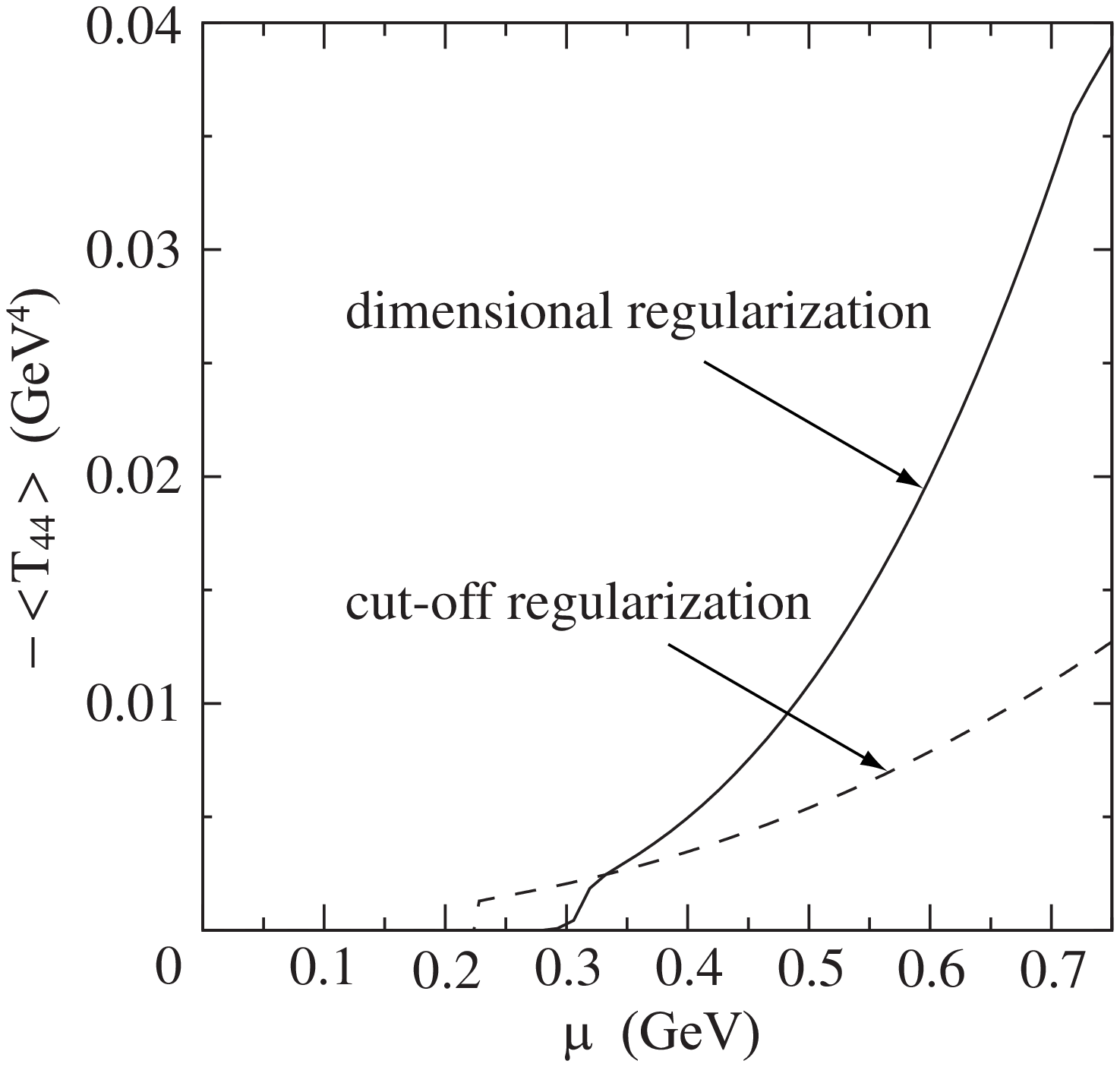}
\caption{Behavior of the stress tensor $-\langle T_{44} \rangle$ 
as functions of the chemical potential at $T=1~{\rm keV}$. 
}
\label{fig_t44}
\end{figure}
\begin{figure}[ht]
 \includegraphics[width=!,height=55mm]
 {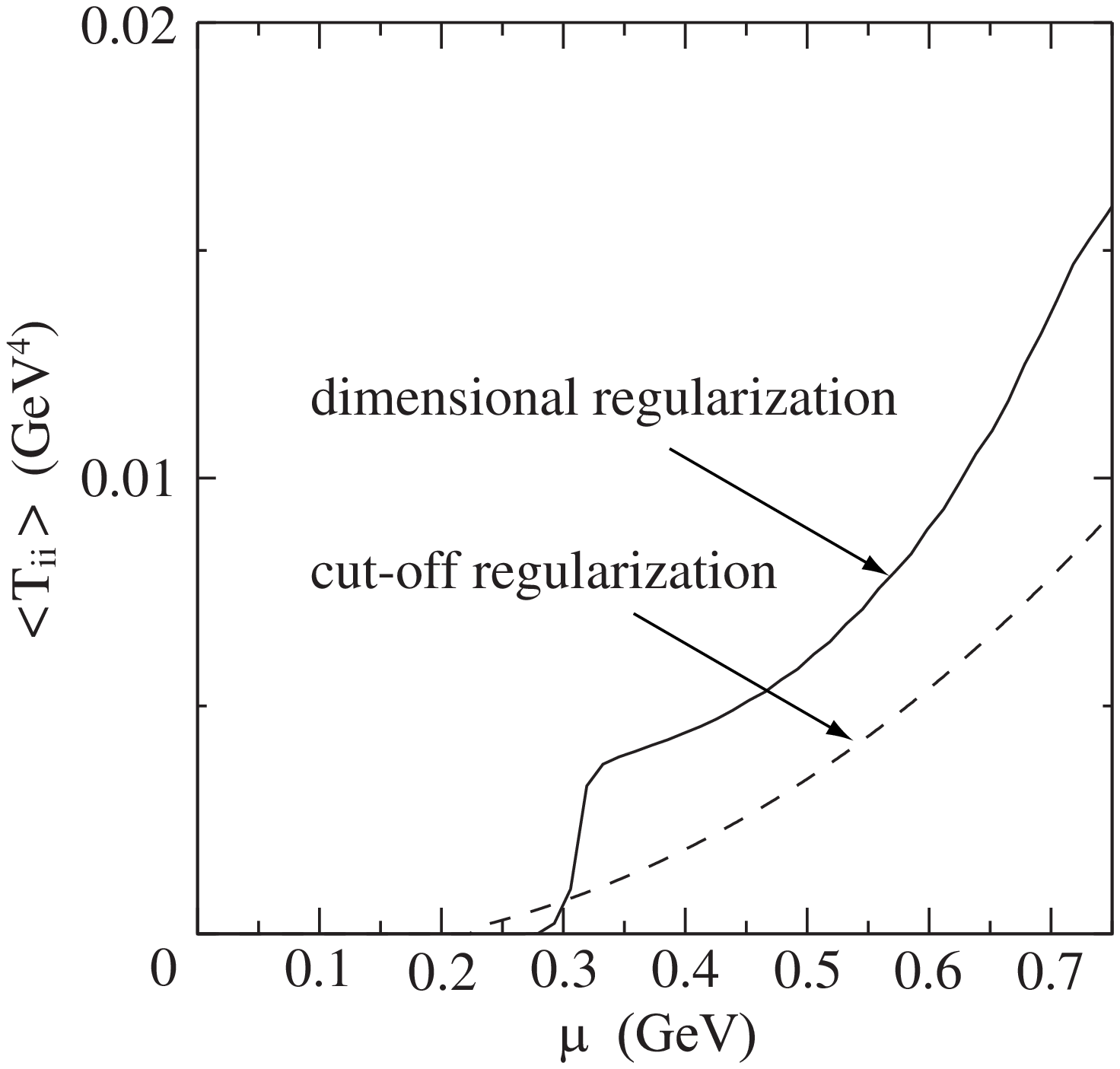}
\caption{Behavior of the stress tensor $\langle T_{ii} \rangle$ 
as functions of the chemical potential at $T=1~{\rm keV}$. 
}
\label{fig_tii}
\end{figure}
\begin{figure}[ht]
 \includegraphics[width=!,height=55mm]
 {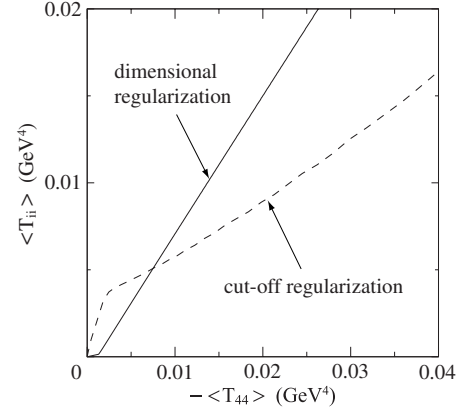}
\caption{Equation of state, i.e.,
relationship between $- \langle T_{44} \rangle$ and 
$\langle T_{ii} \rangle$.
}
\label{fig_t44-tii}
\end{figure}

We plot the energy density, $-\langle T_{44} \rangle$, in 
Fig.~\ref{fig_t44} and the pressure, $\langle T_{ii} \rangle$,
in Fig.~\ref{fig_tii}. 
The origin of the energy density and pressure can not be fixed 
in our model. Here we set $\langle T_{44} \rangle |_{T=\mu=0}=0$
and $\langle T_{ii} \rangle |_{T=\mu=0}=0$ by hand.
In 2SC phase, $\langle \Delta \rangle > 0$, $-\langle T_{44} \rangle$ 
and $\langle T_{ii} \rangle$ monotonically increase as functions of 
$\mu$. We find the equation of state from these results. 
It plays an essential role in the study of the structure of dense stars. 
As is shown
in Fig.~9, there is a difference between the dimensional and 
the cut-off regularizations. A harder state is observed 
in the cut-off regularization for  smaller $-\langle T_{44} \rangle$.
On the other hand, a harder state is obtained in the dimensional
regularization for larger $-\langle T_{44} \rangle$. It affects
the core structure of dense stars with 2SC phase inside.

Next we examine the thermodynamic relationships. In the thermodynamics 
the energy density $\rho$ is derived from the generating functional, $Z$,
through the following differentiation,
\begin{equation}
  \rho\Omega =
  \frac{\partial}{\partial\beta}\left(\ln Z\right)
  -\mu\frac{\partial }{\partial \mu}\frac{\ln Z}{\beta} .
\end{equation}
The pressure of the system, $P$, is obtained by differentiating $Z$ 
 with respect to $\Omega$,
\begin{equation}
  P=\frac{\partial}{\partial \Omega} \frac{\ln Z}{\beta} .
\end{equation}
In the present approximation for the generating functional
the energy density and the pressure become 
\begin{equation}
 \rho =
  \frac{\partial}{\partial\beta}\left(\beta V_{\rm eff}\right)
  -\mu\frac{\partial }{\partial \mu}V_{\rm eff} 
\label{ex:rho}
\end{equation}
and
\begin{equation}
  P = - V_{\rm eff} .
\end{equation}

First we consider the expression for the energy density.
Substituting the explicit expression for the effective potential 
(\ref{veff:ang}) into the first term of the right hand side 
of the Eq.~(\ref{ex:rho}), we get 
\begin{eqnarray}
 \frac{\partial}{\partial \beta}
  (\beta V_{\rm eff})
  &=&
  \frac{\langle\sigma\rangle^2}{4G_S}+\frac{\langle\Delta\rangle^2}{4G_D} 
  -\frac{\tilde{A}}{\Gamma\left(\frac{D-1}{2}\right)}
  \int^\infty_0 dp\ p^{D-2} 
\nonumber \\
  &&\times \left[
   E_+\tanh\left(\frac{\beta E_+}{2}\right) \right. \nonumber \\
  &&\left. +\frac{\xi_+}{2}\tanh\left(\frac{\beta \xi_+}{2}\right)
  \vphantom{\frac{\beta E_+}{2}}+(\mu \to -\mu) 
  \right]. 
\end{eqnarray}
The second term of the right hand side of the Eq.~(\ref{ex:rho})
is the number density (\ref{n_den}), times the chemical potential $\mu$. 
Therefore the energy density $\rho$ is given by 
\begin{eqnarray}
 \rho &=&
  \frac{\langle\sigma\rangle^2}{4G_S}+\frac{\langle\Delta\rangle^2}{4G_D}
 -\frac{\tilde{A}}{\Gamma\left(\frac{D-1}{2}\right)}
  \int^\infty_0 dp\ p^{D-2}
\nonumber \\
 &&\times \left[
   \frac{E_+^2-\mu\xi_+}{E_+}\tanh\left(\frac{\beta E_+}{2}\right)
  \right. \nonumber \\
 &&\left.
   +\frac{E}{2}\tanh\left(\frac{\beta \xi_+}{2}\right)
   +(\mu \to -\mu)
  \right].
\end{eqnarray}
This expression exactly reproduces 
$-\langle T_{44} \rangle$ of the Eq.~(\ref{ex:t44}). 
Thus the time component of the stress tensor, $-\langle T_{44} \rangle$, 
satisfies the thermodynamic relationship for the energy density 
(\ref{ex:rho}). 

For the space component of the stress tensor, $\langle T_{ii} \rangle$, 
the situation is not so simple. After  partial integration over $p$ in 
the effective potential Eq.~(\ref{veff:ang}) or $\langle T_{ii} \rangle$
of the Eq.~(\ref{ex:tii}), we have
\begin{eqnarray}
P &\simeq& \langle T_{ii} \rangle
 +\frac{\tilde{A}}{(D-1)\Gamma\left(\frac{D-1}{2}\right)}
\nonumber \\
 &&\times \left[ p^{D-1} \left( E_+ + E_- -2
   \sqrt{E^2+\langle\Delta\rangle^2} \right)  
   \right]_0^\infty 
\nonumber \\
 &\sim& \langle T_{ii} \rangle
 +\frac{\tilde{A}\ \mu^2 \langle\Delta\rangle^2}
  {(D-1)\Gamma\left(\frac{D-1}{2}\right)}
  \lim_{p\to \infty} p^{D-4} .
\label{v-tii}
\end{eqnarray}
At the limit, $p \to \infty$, $P$ coincides
with $\langle T_{ii} \rangle$ for $D<4$.
In the cut-off regularization  the cut-off scale is used as
the upper limit (instead of $\infty$) in the second term of the right hand side in
Eq.~(\ref{v-tii}). Since the last term has non-vanishing value,
$P$ does not coincide with $\langle T_{ii} \rangle$.

\section{Radius and mass of dense star}
To see the physical implication of the regularization dependence
we evaluate the radius and the mass of dense star in both the 
dimensional and the cut-off regularizations. For this purpose we 
confine ourselves to the case of  a static and spherically symmetric 
star and do not care about the contribution of the strange quark 
and the neutrality of the star.

Here we numerically evaluate the TOV equations in 
four dimensions using the results of the previous sections.
Inside the stars the gravitational force should balance  the 
pressure of the matter. For a static and spherically symmetric star 
this condition is expressed by TOV equation,
\begin{eqnarray}
\frac{d P(r)}{d r}
&=&
  -G
  \frac{\rho(r) + P(r)}
  {r\{r -2GM(r) \}} \nonumber \\
&& \times \{M(r) + 4\pi r^3P(r)\} , 
\label{eq:tov1} \\
 \frac{d M(r)}{dr}
  &=& 4\pi r^2 \rho(r) ,
\label{eq:tov2}
\end{eqnarray}
where $r$ is the radial distance from the center of a star.
The pressure and the energy density are defined by the energy-momentum
tensor $P(r)=\langle T_{ii} \rangle$ and $\rho(r)=-\langle T_{44}\rangle$.
Because of the spherical symmetry the pressure, the energy density 
and the mass function $M(r)$ are given as functions of $r$.

 We assume that the color superconducting phase can be 
realized inside dense stars and solve the 
Eqs.~(\ref{eq:tov1}) and (\ref{eq:tov2}) with account of the
EoS based on the extended NJL model
\footnote{This procedure is valid until the energy density inside the
star is in the deconfinement phase. Here we evaluate the structure of
a quark star in which confinement phase is realized only near the 
surface of the star and ignore the contribution of the confinement 
phase.}.
We set the initial condition at $r=0$ as $M(r=0)=0$ and 
$\rho(r=0)=\rho_c$ and integrate  the TOV equations. 
The energy density and the pressure decrease monotonically from the 
center to the surface. 
In our numerical analysis we define the surface of the quark star at 
the energy density, $\rho\simeq 0.0017 \mbox{GeV}^4$, which corresponds 
to the ordinary energy density inside protons.

In Fig.~10 we plot the mass of a star as a function of the central 
energy density. For lower $\rho_c$ the mass, $M$, increases with the 
central energy density. Increasing $\rho_c$, we observe the maximal 
value for $M$. The mass decreases after passing through the maximal 
values. No stable object can be constructed in this parameter range 
with a negative slope, as is shown by dotted lines in Fig.~10.
The maximal value for the mass in the dimensional regularization is 
a few times as heavy as that in the cut-off regularization.  

\begin{figure}[ht]
 \includegraphics[width=!,height=55mm]{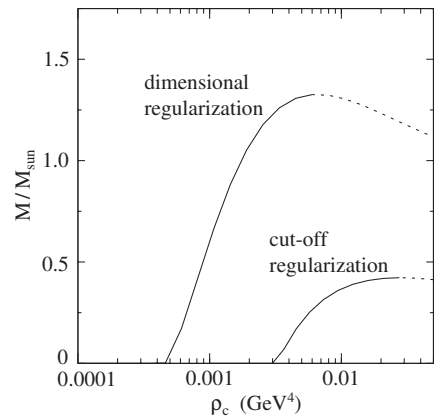}
\caption{Mass of the star as a function of the energy density at the
 center of the star. Solid and dotted lines denote the stable and
 unstable states respectively.}
\label{m-rhoc}
\end{figure}
\begin{figure}[ht]
 \includegraphics[width=!,height=55mm]{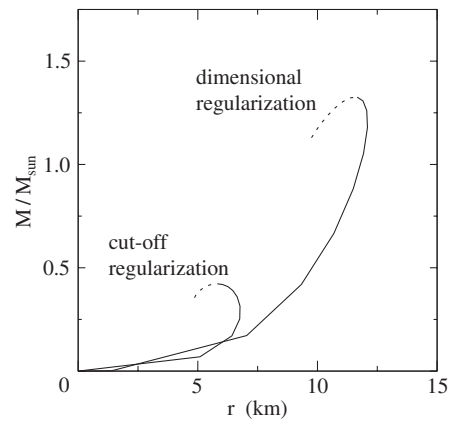}
\caption{Relationship between mass and radius of the star.
Solid and dotted lines denote the stable and unstable states respectively.}
\label{m-r}
\end{figure}

In Fig.~11 the mass of the star is plotted as a function of the 
radius $r$. The dotted lines correspond to the unstable solution 
in Fig.~10. In the dimensional regularization we obtain a heavier and 
a larger quark star. It means that the solution of TOV equations has 
a strong dependence on the regularization scheme.
Since we do not include the contribution from the strange quark,
our result obtained in the cut-off regularization shows
a qualitatively similar behavior but does not exactly reproduce 
the result of the Ref.~\cite{Ruster:2003zh}.

\section{Concluding Remarks}

We have investigated the phase structure of the chiral symmetry 
breaking and the diquark condensation ($SU_c(3)$ symmetry 
spontaneous breaking) on the basis of the 
extended two-flavor 
NJL model by using the dimensional and the cut-off regularizations.

Evaluating the effective potential, we obtain the expectation values 
of $\sigma$ and $\Delta$. The values show a different behavior for 
 the different regularizations.
In the dimensional regularization we obtain a larger mass gap and
no coexistence phase with non-vanishing
$\langle \sigma \rangle$ and $\langle \Delta \rangle$ at the quark 
massless limit, $m \rightarrow 0$.
The diquark condensation, $\langle \Delta \rangle \neq 0$, is caused 
by the Cooper instability of quark pairs near the Fermi surface. 
We loose such a contribution in the cut-off regularization, as the 
chemical potential approaches to the cut-off scale. The diquark 
condensation is suppressed for a larger chemical potential, 
$\mu \sim \Lambda$.
In the dimensional regularization $|\langle \Delta \rangle|$
monotonically increases as the chemical potential increases.

Next we studied the phase structure and the quark number susceptibility.
TCP and CEP for the dimensional regularization are observed at higher 
temperature and lower chemical potential than those for the cut-off 
regularization.
As is seen in Fig.~5, a complex structure is realized around CEP for 
the coexistence phase. 
It is well-known that in the framework of the perturbative QCD the 
2SC is predicted even at asymptotically high values of chemical 
potential $\mu$. However, one of the deficiencies of the NJL model with 
cutoff regularization is the absence of the 2SC phase at sufficiently 
high values of $\mu$. In contrast, it is shown that in the NJL model 
with dimensional regularization the 2SC phase is presented in the phase 
diagram at rather large values of $\mu$, and this conclusion is in 
agreement with perturbative QCD analysis.

The NJL model used in the paper does not describe confinement. 
For the future study of the confinement-deconfinement transitions
we are aware of the papers where  confinement is simulated in NJL
by introduction of the IR cutoff (to lift up the threshold of 
the decay into quarks) \cite{Wolf:01, Wolf:03}.
There is also another way to address confinement \cite{Step:00},   
where they just compare the binding energies per a quark for the case 
of diquark and nucleon clustering.

Confinement may have been relevant to our previous paper
\cite{Inag-08}, where we discussed the width of sigma in the leading 
$1/N$ order. In this paper we claimed that the problem
of the unphysical width can be fixed at the next $1/N$ order, where 
the decay of sigma into $2\pi$ takes place.
So we hoped to fix the problem of the sigma width without even 
simulating confinement.

We have also evaluated the energy-momentum tensor and the EoS. 
The dependence on the regularization observed in the high density region
is not small.
As is seen in Fig.~9, the rate of $\langle T_{ii} \rangle$ and 
$-\langle T_{44} \rangle$ is near unity in the dimensional
regularization. It has been shown that thermodynamic pressure does not 
coincide with the space component of the stress tensor 
$\langle T_{ii} \rangle$ in the cut-off regularization.

Evaluating the TOV equations, we have obtained the relationship between 
the mass and the radius of dense stars in both regularizations.
The existence of heavier and larger quark stars is expected in the 
dimensional regularization.
The regularization dependence is mainly generated by the difference of
the mass gap in Figs.~1 and 2. The diquark condensation 
causes only a little contribution for the mass and radius of dense stars
\cite{Ruster:2003zh}. 
It should be noted that the size of the mass gap depends on the 
regularization parameter $D$. We obtain a smaller mass gap for 
larger dimensions \cite{IKM}.
However, a lower dimensional four-fermion interaction model is 
favored in another approach to QCD phenomena. If we adopt the ladder 
and the 
instantaneous exchange approximations and neglect momentum dependent 
parts of the fermion self-energy, Schwinger-Dyson equation coincides 
with the gap equation in the two dimensional NJL model at the leading order 
of $1/N$ expansion \cite{FIKA08}.

The present work is mainly restricted to the analysis of the regularization 
dependence of observables in the extended NJL model. To apply 
our result to a more realistic case of a quark star  we should 
include the contribution of the strange quarks and take into account the 
electric and the color neutrality. 
It would be also interesting to study the masses of 
$\pi$ and $\sigma$ mesons in the 2SC phase using the dimensional 
regularization and to compare
the results with those in the 2SC phase obtained in the extended 
NJL model with cut-off regularization \cite{Eber:05}.
Currently we are working on these issues. 

\begin{acknowledgments}
The authors appreciate the really useful referees' comments.
T.~Inagaki is supported by the Ministry of Education, Science, 
Sports and Culture, Grant-in-Aid for Scientific Research (C), 
18540276, 2008.
\end{acknowledgments}


\begin{thebibliography}{99}
\bibitem{Bailin:1983bm}
  D.~Bailin and A.~Love, Phys. Rept. {\bf 107}, 325 (1984).
\bibitem{Iwas:1995}
  M.~Iwasaki and T.~Iwado, Phys. Lett. B {\bf 350}, 163 (1995).
\bibitem{Alford:1997zt}
  M.~Alford, K.~Rajagopal and F.~Wilczek, Phys. Lett. B {\bf 422},
  247 (1998).
  M.~Alford, K.~Rajagopal and F.~Wilczek, Nucl. Phys. B {\bf 537}, 443 (1999).
\bibitem{Rapp:1997zu}
  R.~Rapp, T.~Schafer, E.V.~Shuryak and M.~Velkovsky,
  Phys. Rev. Lett. {\bf 81}, 53 (1998).
\bibitem{NJL}
  Y.~Nambu and G.~Jona-Lasinio, Phys. Rev. {\bf 122}, 345 (1960),
  {\it ibid}. {\bf 124}, 246 (1961).
\bibitem{Pro}
  Proc. Int. Workshop on {\it Compact Stars in the QCD Phase Diagram},
  (Copenhagen, Aug. 2001), eConf C010815 (2002);\\
  Proc. ECT Int. Workshop on Physics of Neutron Stars Interiors (NSI00),
  Lect. Notes Phys. 578 (2001).
\bibitem{Alfo:07}
 M.~Alford, D.~Blaschke, A.~Drago, T.~Klahn, G.~Pagliara
 and J.~Schaffner-Bielich, Nature {\bf 445}, E7 (2007). 
\bibitem{Oppenheimer:1939ne}
  J.~R.~Oppenheimer and G.~M.~Volkoff,
Phys.\ Rev.\  {\bf 55}, 374 (1939).
\bibitem{PPLS} D.~Page, M.~Prakash, J.M.~Lattimer, and A.W.~Steiner, 
 Phys. Rev. Lett. {\bf 85}, 2048 (2000).
\bibitem{YIM}
K.~Yamaguchi, M.~Iwasaki and O.~Miyamura, 
Prog. Theor. Phys. {\bf 107}, 117 (2002).
\bibitem{Alford:2002kj}
  M.~Alford and K.~Rajagopal, JHEP {\bf 0206}, 031 (2002).
\bibitem{SRP}
A.W.~Steiner, S.~Reddy, and M.~Prakash, Phys. Rev. D {\bf 66}, 094007 (2002).
\bibitem{Shovkovy:2003ps}
I.~Shovkovy, M.~Hanauske, and M.~Huang, Phys. Rev. D {\bf 67}, 103004
	(2003).
\bibitem{BB}
S.~Banik and D.~Bandyopadhyay, Phys. Rev. D {\bf 67}, 123003 (2003).
\bibitem{Blas:08}
D.~Blaschke, F.~Sandin, T.~Klahn and J. Berdermann, arXiv:0807.0414 [nucl-th].
\bibitem{BHIT}
W.~Bentz, T.~Horikawa, N.~Ishii and A.W.~Thomas, 
 Nucl. Phys. A {\bf 720}, 95 (2003).
\bibitem{LBT}
S.~Lawley, W.~Bentz and A.W.~Thomas, 
 Nucl. Phys. Proc. Suppl. {\bf 141}, 29 (2005).
\bibitem{WLLTW}
P.~Wang, S.~Lawley, D.B.~Leinweber, A.W.~Thomas and A.G.~Williams, 
 Phys. Rev. C {\bf 72}, 045801 (2005).
\bibitem{Blas:07}
T.~Klahn, D.~Blaschke, F.~ Sandin, C.~Fuchs, A.~Faessler, H.~Grigorian,
G.~Ropke and J.~Trumper,  Phys. Lett. B {\bf 654}, 170 (2007).
D.B.~Blaschke, D.~GomezDumm, A.G.~Grunfeld, T.~Klahn and N.N. Scoccola, 
Phys. Rev. C {\bf 75}, 065804 (2007). 
D.B.~Blaschke, T.~Klahn, and F.~Sandin,
J. Phys. G {\bf 35}, 014051 (2008). 
\bibitem{Panda:2006}
P.K.~Panda and H.S.~Nataraj, Phys. Rev. C {\bf 73}, 025807 (2006).
\bibitem{BO}
M.~Buballa and M.~Oertel, Nucl. Phys. A {\bf 703}, 770 (2002).
\bibitem{BBBNO}
M.~Baldo, M.~Buballa, F.~Burgio, F.~Neumann, M.~Oertel, and
H.J.~Schulze, Phys. Lett. B {\bf 562}, 153 (2003).
\bibitem{HLFP}
J.E.~Horvath, G.~Lugones, and J.A.~ de Freitas Pacheco, 
 Int. J. Mod. Phys. D {\bf 12}, 519 (2003).
\bibitem{LH}
G.~Lugones and J.E.~Horvath, Astron. Astrophys. {\bf 403}, 173 (2003).
\bibitem{AR}
M.~Alford and S.~Reddy, Phys. Rev. D {\bf 67}, 074024 (2003).
\bibitem{GBA}
H.~Grigorian, D.~Blaschke, and D.N.~Aguilera,
 Phys. Rev. C {\bf 69}, 065802 (2004). 
\bibitem{BGAYT}
D.~Blaschke, H.~Grigorian, D.N.~Aguilera, S.~Yasui, and H.~Toki, 
AIP Conf. Proc. Acoust. Phys. {\bf 660}, 209 (2003).
\bibitem{Ruster:2003zh}
  S.B.~Ruster and D.H.~Rischke, Phys. Rev. D {\bf 69}, 045011 (2004)
\bibitem{BFG}
D.~Blaschke, S.~Fredriksson, H.~Grigorian, A.M.~Oztas
and F.~Sandin, Phys. Rev. D {\bf 72}, 065020 (2005).
\bibitem{Hatt:03}
  Y.~Hatta and T.~Ikeda,
   Phys. Rev. D {\bf 67}, 014028 (2003). 
\bibitem{Scha:07}
  B.-J.~Schaefer and J.~Wambach
   Phys. Rev. D {\bf 75}, 085015 (2007). 
\bibitem{Cost:07}
  P~Costa, C.A.~de~Sousa, M.C.~Ruivo and Yu.L.~Kalinovsky
   Phys. Lett. B {\bf 647}, 431 (2007). 
  P.~Costa, M.C.~Ruivo and C.A.~de~Sousa,
   Phys. Rev. D {\bf 77}, 096001 (2008). 
\bibitem{Sasa:08}
  C.~Sasaki, B.~Friman and K.~Redlich
   Phys. Rev. D {\bf 77}, 034024 (2008). 
\bibitem{Fuku:08}
  K.~Fukushima,
   Phys. Rev. D {\bf 77}, 114028 (2008). 
\bibitem{NG}
L.~P.~Gorkov, JETP {\bf 7}, 993 (1958),
Y.~Nambu,
Phys.\ Rev.\  {\bf 117}, 648 (1960).
\bibitem{Fuji-07}
  T.~Fujihara, T.~Inagaki and D.~Kimura, 
  Prog. Theor. Phys. {\bf 117}, 139 (2007). 
\bibitem{IKM}
  T.~Inagaki, T.~Kouno and T.~Muta,
  Int. J. Mod. Phys. A {\bf 10}, 2241 (1995).
\bibitem{Inag-08}
  T.~Inagaki, D.~Kimura and A.~Kvinikhidze,
  Phys. Rev. D {\bf 77}, 116004 (2008).
\bibitem{Naka:92}
  S.~Krewald and K.~Nakayama,
   Annals Phys. {\bf 216}, 201 (1992).
\bibitem{Jafa:06}
  R.G.~Jafarov, and V.E.~Rochev,
   Russ. Phys. J. {\bf 49}, 364 (2006), 
   Izv. Vuz. Fiz. {\bf 49}, 20 (2006). 
\bibitem{KY}
  Y.~Kikukawa and K.~Yamawaki, Phys. Lett. B {\bf 234}, 497 (1990).
\bibitem{M}
  T.~Muta, {\it Nagoya Spring School on Dynamical Symmetry Breaking}, 3
  ( ed. K.~Yamawaki, World Scientific, 1992).
\bibitem{HKWY}
  H.-J.~He, Y.-P.~Kuang, Q.~Wang and Y.-P.~Yi, 
  Phys. Rev. D {\bf 45}, 4610 (1992).
\bibitem{ps1}
  P.~A.~M.~Dirac, Proc. Cambridge Phil. Soc. {\bf 30}, 150 (1934).
\bibitem{ps2}
  R.~Peierls, Proc. Roy. Soc., Series A {\bf 146}, 420 (1934).
\bibitem{ps3}
  J.~Schwinger, Phys. Rev. {\bf 82}, 664 (1951).
\bibitem{P}
  N.~Petropoulos,
  J. Phys. G {\bf 25}, 225 (1999).
\bibitem{LR}
  J.~T.~Lenaghan and D.~H.~Rischke,
  J. Phys. G {\bf 26}, 431 (2000).
\bibitem{Wolf:01}
  W.~Bentz and A.W.~Thomas,
   Nucl. Phys. A {\bf 696}, 138 (2001).
\bibitem{Wolf:03}
  W.~Bentz, T.~Horikawa, N.~Ishii and A.W.~Thomas
   Nucl. Phys. A {\bf 720}, 95 (2003). 
\bibitem{Step:00}
  S.~Pepin, M.C.~Birse, J.A.~McGovern and N.R.~Walet, 
   Phys. Rev. C {\bf 61}, 055209 (2000).
\bibitem{FIKA08}
  T.~Fujihara, T.~Inagaki, D.~Kimura and A.~Kvinikhidze,
  Prog. Theor. Phys. Suppl. {\bf 174}, 72 (2008):
  arXiv:0806.1331 [hep-ph]
\bibitem{Eber:05}
  D.~Ebert, K.G.~Klimenko and V.L.~Yudichev,
   Phys. Rev. C {\bf 72}, 015201 (2005),
   Phys. Rev. D {\bf 72}, 056007 (2005), 
   {\it ibid}. {\bf 75}, 025024 (2007).
\end{thebibliography}
\end{document}